\documentstyle[aps, psfig]{revtex}

\begin{document}
\title{Excitation spectrum of vortex lattices in rotating Bose-Einstein condensates}
\author{S. Choi,  L. O. Baksmaty, S. J. Woo, and N. P. Bigelow}
\address{Department of Physics and Astronomy, University of Rochester, Rochester, NY 
14627}
\maketitle

\begin{abstract}
Using the coarse grain averaged hydrodynamic approach, we calculate the excitation 
spectrum of vortex lattices sustained in rotating Bose-Einstein condensates. The spectrum 
gives the 
frequencies of the common-mode longitudinal waves in the hydrodynamic regime, 
including those of the higher-order compressional modes. Reasonable agreement with the 
measurements taken in a recent JILA experiment is found, suggesting that one of the 
longitudinal modes reported in the experiment is likely to be the $n=2, m=0$ mode.  
\end{abstract}


\vspace{6mm}

Quantized vortices constitute one of the major topics in the study of Bose-Einstein 
Condensates (BEC)\cite{FetterReview}. Following the demonstration of the nucleation of 
quantized vortices in BEC\cite{Matthews,Foot}, vortex lattices have since been produced 
experimentally\cite{Dalibard,Ketterle1,Ketterle2,Cornell}, and very recently, both the 
transverse (Tkachenko)
and the logitudinal (hydrodynamic) excitations of the vortex lattices have been observed at 
JILA\cite{Cornell2}.

Related systems of vortex lattices in type-II superconductors and in superfluid $^4$He  
have been studied extensively in the past, resulting in a large number of important 
theoretical studies\cite{Tkachenko,CampbellZiff,Campbell,Sonin}. 
Theoretical investigations of the new BEC vortex lattices
have also been gaining 
momentum\cite{CastinDum,ButtsRokhsar,Tsubota,SinatraCastin,FetterLattice,Kasamatsu}.
In particular, Anglin and Crescimanno have provided an extension to the 
Feynman-Tkachenko continuum theory for a dense vortex lattice and thereby studied the 
long wavelength 
transverse excitations at slow rotation taking full account of the finite geometry\cite{Anglin}, while 
Baym\cite{Baym} has calculated the frequencies of Tkachenko waves at general rotation speeds from Thomas-Fermi to 
the mean-field quantum Hall regimes, with the calculated frequencies  in  
good agreement with the observations of Tkachenko modes at JILA. 
In addition, Cozzini, Chevy and Stringari have studied the hydrodynamic  excitations of the surface, 
scissors and Kelvin  modes\cite{StringariLattice}.

In this paper we calculate the frequencies of the logitudinal excitations of vortex lattices by 
following the procedure used by Svidzinsky and Fetter to study low-lying modes 
of a single vortex line in BEC\cite{FetterNormalMode}.  
This enables us to write the excitation frequencies in terms of the frequencies of a 
vortex-free condensate, and as a corollary we are able to estimate the frequencies of the 
higher-order compressional modes as well as the surface excitations.

Although a vortex lattice is significantly more complicated than a single vortex, the concept 
of diffused vorticity, ${\bf \nabla} \times {\bf  v} = 2 {\bf \Omega}$ of a vortex lattice 
provides an effective coarse grain averaged description that greatly simplifies the 
formalism by ``smoothing out'' the effect of individual vortices\cite{Sonin}.    
The density of vortex lines per unit surface is $n_{\nu} = 2 \Omega/k$ where $\Omega$ is 
the angular velocity of the
containing vessel and $k = h/M$ is the circulation. From the area per vortex, $n_{\nu}^{-
1} = \pi b^2$, one may define the intervortex separation $b = \sqrt{\hbar/M \Omega}$, 
which provides, along with the healing length,  a characteristic length  scale for the vortex 
lattice. The healing length $\xi = 1/\sqrt{8 \pi n a}$, where $n$ is the density of the gas and 
$a$ is the $s$-wave scattering length, provides an estimate of the  size of the vortex core.  
The functional dependence of $n_{\nu}$  on $\Omega$  
implies that the vortex lines rotate with the containing vessel; it has indeed been observed in
superfluid $^{4}$He and in alkali BEC's that vortex lattices undergo self-consistent solid-
body
rotation.

A description of the dynamics of a system containing a vortex lattice can therefore be
more conveniently given by transforming to the frame rotating with angular
velocity $\Omega$. This renders the Hamiltonian time-independent and hence facilitates 
determination of the metastable (rotating) ground state. The free energy in the rotating 
frame can be written in the following form\cite{FetterLattice}: 
\begin{equation}
F = \int d{V} \left [ \frac{\hbar^2}{2M} ({\bf \nabla} |\Psi| )^{2} + V(r)
|\Psi|^{2} + \frac{1}{2} g |\Psi|^{4} + \frac{1}{2} M ({\bf v} - {\bf v}%
_{sb})^{2} |\Psi|^{2} - \frac{1}{2} M \Omega^{2} {r}^{2} |\Psi|^{2} \right ],
\label{freeenergy}
\end{equation}
where $V({r})$ is the confining potential and $g = 4 \pi a \hbar^2 /M$ is
the interparticle coupling constant with the $s$-wave scattering length $a$. 
${\bf v}_{sb}$ denotes the velocity of solid body rotation, ${\bf v}_{sb} = 
{\bf \Omega} \times {\bf r}$, such that ${\bf v}-{\bf v}_{sb}$ defines the
velocity of the fluid in the rotating frame, and $- \frac{1}{2} M \Omega^{2} 
{r}^{2} $ represents an effective repulsive centrifugal potential.

From the Gross-Pitaevskii Equation (GPE) associated with the free energy,
one can write the Thomas-Fermi (TF) profile for the vortex lattice as
$|\Psi_{TF}({r})|^2 \approx |\Psi^{(0)}_{TF}(r)|^2 u^2(r)$ where $|\Psi^{(0)}_{TF}|^2$  
is the TF density in the absence of a vortex lattice 
\begin{equation}
|\Psi^{(0)}_{TF}({r})|^2 = \frac{\mu_{\Omega}}{g} \left ( 1 - \frac{M
\omega_{trap}^2 {\ r}^2}{2\mu_{\Omega}} + \frac{M \Omega^2 {\ r}^2}{%
2\mu_{\Omega}} \right ),
\end{equation}
and 
\begin{equation}
u^2(r) = 1 - \frac{M ({\bf v} - {\bf v}_{sb})^{2}}{2\mu_{\Omega}}
\end{equation}
provides the periodic vortex lattice structure\cite{FetterLattice}.  
In any given unit cell of the vortex lattice, assuming a circular Wigner-Seitz cell,  $u^2(r)$ 
has the local form 
\begin{equation}
u^2({\ r}) = 1 - \frac{\xi^2}{{\ r}^2} + \frac{2\xi^2}{b^2} - \frac{\xi^2 {\
r}^2}{b^4},
\end{equation}
where $\xi \leq r \leq b$\cite{FetterLattice,Fetter83}.
In the dense vortex limit $\xi \ll b$, i.e. the case of large condensate with many vortices
for which the concept of diffused vorticity works well, one may approximate $\langle
u^2 \rangle \approx 1$. We therefore write the TF profile as 
\begin{equation}
|\Psi_{TF}({r})|^2 = \frac{\mu_{\Omega}}{g} \left ( 1 - \frac{M
(\omega_{trap}^2 - \Omega^2) {\ r}^2}{2\mu_{\Omega}} \right ) .
\label{TFdensity}
\end{equation}
It is noted that the net effect of a vortex lattice is as if the trapping
potential has been modified by that with an $\Omega$-dependent spring
constant. With the self-consistent rotation in place, a 3-dimensional (3D) condensate 
essentially takes on a pancake shape, closely approximating a 2D system\cite{Cornell2}; 
we shall  therefore consider 2D BEC in this paper. 
From the usual considerations of normalization, the chemical
potential of a vortex lattice in 2D is
\begin{equation}
\mu_{\Omega} = \mu_0 \sqrt{1 - \frac{\Omega^2}{(\omega_{trap})^2}}.
\end{equation}

Using the GPE under the zero temperature approximation, i.e. almost all particles are 
assumed to be 
in the condensate, one can write the corresponding hydrodynamic equations of motion
with the substitution $\Psi = n \exp [iS({\bf r})]$ where $n$ is the 
condensate particle density, $n = |\Psi|^2$, and $S$ is the condensate phase from which the 
condensate velocity is defined: ${\bf v} = \hbar {\bf \nabla} S({\bf r}) /M$. The resulting 
hydrodynamic equations in the TF approximation in which the quantum pressure term of 
the form 
$(\hbar^2/2m \sqrt{n}) \nabla^2 \sqrt{n}$  is ignored 
 are the well-known continuity and Euler equations of fluid dynamics\cite{Stringari}. 
When transformed to the 
rotating frame, the equations take the form\cite{Sonin}: 
\begin{equation}
\frac{\partial n}{\partial t} + {\bf \nabla} \cdot (n {\bf v}) = 0,
\label{continuity}
\end{equation}
\begin{equation}
M\frac{\partial {\bf v}}{\partial t} +  {\bf \nabla}   \left ( \frac{1}{2}Mv^2
+  V_{trap}({\bf r}) + g n \right ) = M{\bf v}
\times (2{\bf \Omega} + {\bf \nabla} \times {\bf v}) + {\bf \nabla} \frac{1}{2} 
M\Omega^2 {\bf r}^2 ,
\label{euler}
\end{equation}
where, on the right hand side, the terms $M{\bf v} \times 2{\bf \Omega}$ and ${\bf 
\nabla} 
\frac{1}{2} M \Omega^2 {\bf r}^2$ provide, respectively, the Coriolis and the centrifugal 
forces experienced by the fluid. These terms vanish in the non-rotating frame.  In order to 
solve the equations, we expand the velocity field in the rotating frame, ${\bf v}$, 
and the  density $n$ to first order in fluctuations.

Writing $n= n_0 + n^{\prime}$ and ${\bf v} = {\bf v}_0 + {\bf v}^{\prime}$
where $n_0$ and ${\bf v}_0$ are the equilibrium values, we linearize these
hydrodynamic equations to determine the harmonic fluctuations in the
particle density $n^{\prime}e^{-i\omega t}$ and the velocity ${\bf v}' e^{-i\omega t}$. 
Following the steps in Ref. \cite{Stringari,Fetter96} but
for the rotating frame, the equations obeyed by $n^{\prime}$ and ${\bf v}^{\prime}$ 
are found to be
\begin{eqnarray}
i \omega n^{\prime}& = & {\bf \nabla} \cdot ( n_0 {\bf v}^{\prime}), \label{nprime} \\
i \omega {\bf v}^{\prime}  & =&  \frac{g}{M} {\bf \nabla} n^{\prime} + 2 {\bf \Omega} 
\times {\bf v}' \label{vprime} ,  
\end{eqnarray}
where we have made the continuous vorticity approximation for the velocity field in the 
rotating frame:
\begin{equation}
{\bf v}_0 \approx 0 \; {;} \;\;\; {\bf \nabla} \times  {\bf v}' \approx 0.  \label{rotapprox}
\end{equation}
This is equivalent to assuming  diffused vorticity for the velocity field in the non-rotating 
frame,
${\nabla} \times {\bf v} \approx 2 {\bf 
\Omega}$. Equation (\ref{rotapprox}) is consistent with the approximation $\xi \ll b$,  since the 
deviation of the velocity field from that of the solid body rotation occurs mainly near vortex cores.  
 It should also be noted that, if the number of particles 
$N$ is small, the contribution of the quantum pressure term may be neglected safely only 
for the lowest multipoles.

Equations (\ref{nprime}-\ref{vprime}) may be written in terms of the velocity components 
${v}^{\prime}_{r}$ and ${v}^{\prime}_{\phi}$ of  ${\bf v}^{\prime} \equiv v'_{r}
\vec{r} + v'_{\phi} \vec{\phi}$  as follows:
\begin{eqnarray}
i \omega n^{\prime}& = & \frac{\partial  n_0}{\partial r} v'_{r} + n_0  \left (\frac{\partial 
v'_{r}}{\partial r} + \frac{1}{r}\frac{\partial v'_{\phi}}{\partial \phi} \right ), \\
i \omega {v}^{\prime}_{r}  & = & \frac{g}{M} \frac{\partial n^{\prime}}{\partial r} -  
2\Omega  v'_{\phi} ,  \label{vr} \\ 
i \omega {v}^{\prime}_{\phi}  & =& \frac{g}{M} \frac{1}{r}\frac{\partial 
n^{\prime}}{\partial \phi} +  2\Omega v'_{r}  \label{vphi}.
\end{eqnarray}

We introduce the velocity potential $\Phi^{\prime}$ where ${\bf 
v}^{\prime}= {\bf \nabla} \Phi^{\prime}$, and  
assume that, in our coarse grained description, the normal modes are proportional to 
$e^{im\phi}$ i.e. 
\begin{equation}
n^{\prime}\propto \exp(im\phi) \; {;}  \;\;\;  \Phi^{\prime}\propto \exp(im\phi).  
\label{phidependence}
\end{equation}
Using Eq. (\ref{phidependence}), and integrating Eq. (\ref{vr}) with respect to $r$,  we 
get
\begin{eqnarray}
i \omega n^{\prime}& = &  \frac{1}{r} \frac{\partial }{\partial r} \left ( r n_0
 \frac{\partial \Phi^{\prime}}{\partial r} \right ) + \frac{n_0 m^2}{r^2} \Phi^{\prime} ,
\label{nprime2} \\
i \omega \Phi'  & = & \frac{g}{M}  n^{\prime} -  2\Omega   i m \int \frac{1}{r} \Phi' dr ,  
\label{vr2} \\ 
i \omega  \Phi' & = & \frac{g}{M} n' +  2\Omega \frac{r}{im} \frac{\partial 
\Phi'}{\partial r}.  \label{vphi2}
\end{eqnarray}

Equating the time derivatives of the velocity potential $\Phi'$, from Eqs. (\ref{vr2}) and 
(\ref{vphi2}), one gets 
\begin{equation}
\frac{r}{m^2}  \frac{\partial \Phi'}{\partial r} =  \int \frac{1}{r} \Phi' dr , \label{aux} 
\end{equation}
which can be re-written as a Cauchy equation for the radial component of $\Phi'$
\begin{equation}
r^2 \frac{\partial^2 \Phi'}{\partial r^2} + r \frac{\partial \Phi'}{\partial r} - m^2 \Phi' = 0    
\label{cauchy} 
\end{equation}
with a solution $\Phi' =  c_1 \exp(m \ln r) +  c_2 \exp(-m \ln r)$,
where $c_1$ and $c_2$ are arbitrary constants. The derivative with respect to $r$ of 
$\Phi'$ gives the velocity in the radial direction
\begin{equation}
\frac{\partial \Phi'}{\partial r}  =  \frac{m}{r} \left [ c_1 \exp(m \ln r) -  c_2 \exp(-m \ln r) 
\right ] ,
\end{equation}  
and since the radial velocity of the excitation cannot diverge as $r \rightarrow 0$, $c_2 = 0$ 
for 
$m > 0$, and $c_1 = 0$ for $m < 0$. Changing notations such that $c_1 \equiv c^{(+)}$ 
and 
$c_2 \equiv c^{(-)}$, one may write 
\begin{equation}
\frac{\partial \Phi'}{\partial r}  = \pm \frac{ m}{r}  c^{(\pm)} \exp(|m| \ln r) = \pm  \frac{ 
m}{r}  \Phi'
\end{equation}  
where we have explicitly specified the existence of the positive and negative $m$ values. 
The equation obeyed by the velocity potential $\Phi'$ consistent with the assumptions that 
we have made is finally:
\begin{equation}
i \omega  \Phi'  =  \frac{g}{M} n' \mp 2 i \Omega \Phi' . \label{phi_r} \\
\end{equation}
Rearranging Eq. (\ref{phi_r}) and substituting in Eq. (\ref{nprime2}), one obtains
\begin{equation}
\frac{g}{M} \left [ \frac{1}{r} \frac{\partial }{\partial r} \left ( r
n_0 \frac{\partial n^{\prime}}{\partial r} \right ) - \frac{m^2%
}{r^2} n_0 n^{\prime}\right ]  +  \omega (\omega \pm 2 \Omega )  n^{\prime} = 0 .  
\label{almostfinal}
\end{equation}
We now introduce the dimensionless units: $\omega^{\prime}= \omega_{trap}$, $%
\Omega^{\prime}= \Omega/\omega_{trap}$, $r^{\prime}= r/R_{\Omega}$ where $%
R_{\Omega}^2 = 2 \mu_{\Omega} / M (\omega^2_{trap} - \Omega^2)$. Making the
substitutions in Eq. (\ref{almostfinal}), we get
\begin{equation}
\frac{(1 - \Omega^{^{\prime}2})}{2} \left [ (1 - r^{^{\prime}2} ) \frac{
\partial^2 n^{\prime}}{\partial r^{\prime 2}}  + \left ( \frac{1}{r^{\prime} }  -   
3r^{\prime} \right ) \frac{\partial
n^{\prime}}{\partial r^{\prime}} - \frac{m^2}{%
r^{^{\prime}2}} (1 - r^{^{\prime}2}) n^{\prime} \right ]  
 +   \omega' (\omega' \pm 2 \Omega')  n^{\prime}= 0.
\label{finalvortex}
\end{equation}

In order to calculate the eigenvalues, the density fluctuation $n^{\prime}$
needs to be specified. The functional form for the excitation of the
vortex-free condensate is $n_0^{\prime \prime}\propto
r^{^{\prime\prime}|m|}P_{n}^{(|m|,0)}(1-2r^{^{\prime\prime}2})$ where $%
P_{n}^{(a,b)}(x)$ is a Jacobi polynomial of order $n$ and the scaled
position variable $r^{\prime\prime}= r/R_0$, with $R_0 = \sqrt{2\mu_0/M
\omega^2_{trap}}$\cite{Fliesser,Ohberg}. Since $n^{\prime}$ must tend to $%
n_0^{\prime \prime}$ as $\Omega^{\prime}\rightarrow
0 $, i.e. the wave function $n'$ should tend to that for the vortex-free case
as the rotational frequency tends to zero, and assuming the net effect 
of a coarse grain averaged vortex lattice on the wave functions to be 
a modification of the spring constant for the trapping potential similar to Eq. 
(\ref{TFdensity}) for the TF ground state, we 
shall use as our ansatz that $n^{\prime}(r^{\prime}) \propto
r^{^{\prime}|m|}P_{n}^{(|m|,0)}(1-2r^{^{\prime}2})$.

The equation obeyed by $n_0^{\prime \prime}$ is known to be 
\begin{equation}
\frac{1}{2} \left [ (1 - r^{^{\prime\prime}2} ) \frac{\partial^2
n_{0}^{\prime \prime}}{\partial r^{^{\prime\prime}2}} + \left ( \frac{1}{%
r^{^{\prime\prime}}} - 3r^{^{\prime\prime}} \right ) \frac{\partial
n_{0}^{\prime \prime }}{\partial r^{\prime\prime}} - \frac{m^2 
}{r^{^{\prime\prime}2}} (1 -
r^{^{\prime\prime}2}) n_{0}^{\prime \prime}\right ] + (\omega^0)^2 n_{0}^{\prime 
\prime}= 0,
\end{equation}
with the eigenvalues given by $(\omega^0_{mn})^2 = |m| + 2n(n+|m| + 1)$\cite
{Fliesser,Ohberg}. With our ansatz for $n^{\prime}(r^{\prime})$, the
expectation value of Eq. (\ref{finalvortex}) implies
\begin{equation}
\omega' (\omega' \pm 2 \Omega')   = \left [ |m| + 2n(n+|m| + 1) \right ] (1 -
\Omega^{^{\prime}2}),
\end{equation}
which may be solved to give the eigenfrequencies for the vortex lattice of a 2D BEC in the 
rotating frame:
\begin{equation}
\omega'_{n, \pm m} = \sqrt{ \left [ |m|+ 2n(n+|m| + 1) \right ] (1 - \Omega^{^{\prime}2}) 
+ \Omega'^2 } \mp \Omega'.
\end{equation}
The solution indicates that energy degeneracy is removed for all $m$ in the rotating frame, 
with the frequency separation of $2\Omega'$. 
By calculating the spectrum in the rotating frame, the excitation energy is given relative to  
the metastable ground state of rotating vortex lattice. As to be expected, the spectrum 
reduces to the well-known result for a vortex-free 
BEC as $\Omega' \rightarrow 0$.  One 
may also proceed in the lab frame for comparison with experiments; the
corresponding rotational hydrodynamic equations in the lab frame are given by Eqs. 
(\ref{continuity}-\ref{euler}) but with the right hand side of Eq. (\ref{euler}) having only 
the $M {\bf v} \times ({\bf \nabla} \times {\bf v})$ term, and ${\bf v}$ denoting the 
velocity field in the lab frame.  
Following through with the calculation one obtains the spectrum in the lab frame:
\begin{equation}
\omega'_{n, \pm m} = \sqrt{ \left [ |m|+ 2n(n+|m| + 1) \right ] (1 - \Omega^{^{\prime}2}) 
+ \Omega'^2 } \pm  (m - 1) \Omega'. \label{spectrum}
\end{equation}
This is as  expected since the free energy in the rotating frame, 
$F$, is related to the energy in the lab frame, $H$, via $F = H - L_z\Omega$ 
where $L_z = {\bf r} \times {\bf p}$ is the $z$-component angular momentum operator. 
The frequency of the $m=0$ modes is identical in both the rotating and non-
rotating frames, and energy degeneracy is restored for the $m = \pm 1$ modes in the lab 
frame. It should be noted that, due to the nature of the 
hydrodynamic calculation, the frequencies predicted by Eq. (\ref{spectrum}) correspond to 
those of the common-mode logitudinal wave of the vortex lattice rather than the transverse 
Tkachenko waves. It is straightforward to see from Eq. (\ref{spectrum}) that one can write 
for the surface ($n=0$) modes $\omega^{\prime}_{0,\pm m} = \sqrt{|m| - (|m| - 1) 
\Omega^{\prime 2} } \pm  
(m - 1) \Omega'$, in agreement with the result of Ref. \cite{StringariLattice}. 
In particular, the dipole (or the center of mass mode gives $\omega'_{0,\pm 1} = 1$ for all 
$\Omega'$, and the quadrupole mode gives $\omega'_{0,\pm 2} = \sqrt{2 - \Omega'^2} 
\pm \Omega'$ which accurately predicts the experimental value of the quadrupole 
($n=0,m=\pm 2$) excitations\cite{Cornell}.

In Ref. \cite{Cornell2}, experimental measurement of the frequencies of the common-mode 
longitudinal waves have been reported for the condensate rotating at $\Omega'= 0.95$. Of 
the three distinct $m=0$ modes that have been observed, the nature of the two of the modes 
have been identified while the third mode could not be classified conclusively. The two 
modes 
that could be identified were the Tkachenko s-bend mode at $0.072 
\frac{\omega_{t}}{2\pi}$ and a radial breathing mode at $2.0 \pm 0.1 
\frac{\omega_{t}}{2\pi}$. The third mode, which was presumed likely to be another 
logitudinal mode, was found to occur at $2.2 \pm 0.1 \frac{\omega_{t}}{2\pi}$. For 
$\Omega'= 0.95$, one of the frequencies 
predicted by Eq. (\ref{spectrum}) for the $n = 1,m=0$ radial breathing mode is 
$\omega'_{1,0} = 2.09$  while the frequency for the $n = 2,m=0$  mode is predicted to 
be
$\omega'_{2,0} = 2.39$. These values differ from the reported experimental values by 5 
and 7\% respectively,  or even less when the experimental error bars are taken into account. 
We believe therefore that the third $m=0$ mode observed in Ref. \cite{Cornell2} is  
the $n=2, m=0$ compressional mode.

\begin{figure}[t]
\centerline{\psfig{height=11cm,file=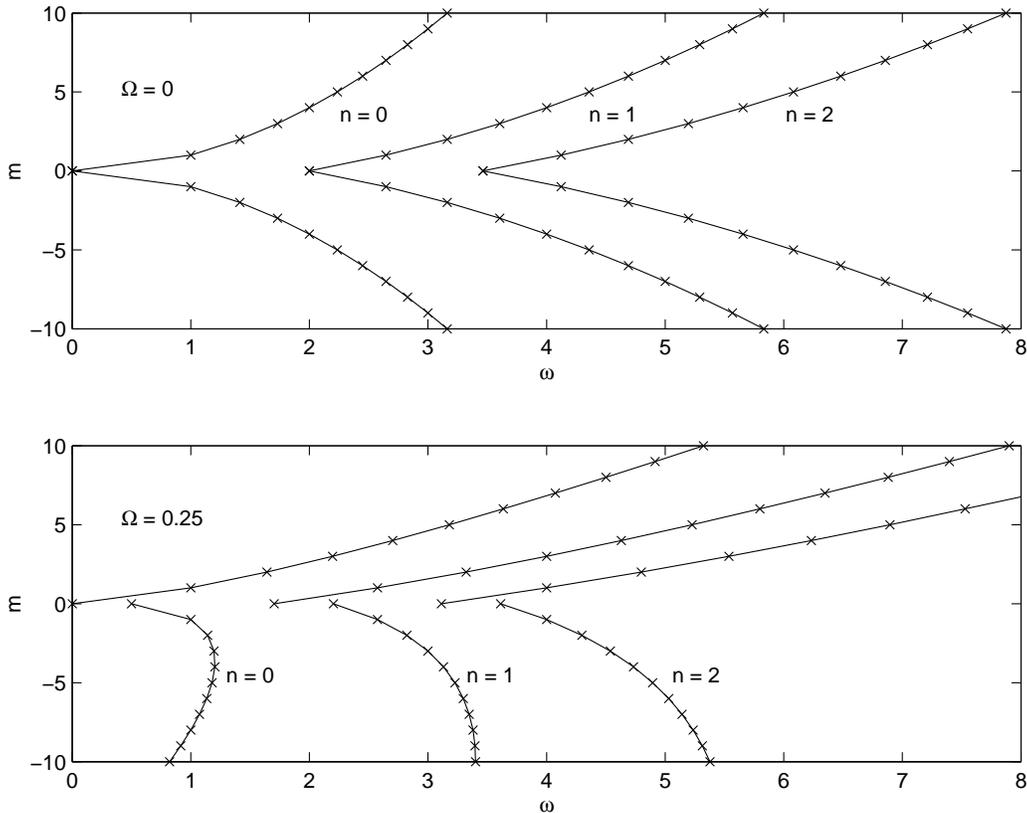}}
\caption{Excitation spectrum, $\omega$ vs. $m$. Points with the same radial quantum 
numbers $n$ are joined by a solid line. The top graph is for $\Omega'= 0$ i.e. in the 
absense of rotation and hence no vortex lattice; the bottom graph is for a rotating 
condensate with $\Omega'= 0.25$.} \label{spectrumfig}
\end{figure}

\begin{figure}[t]
\centerline{\psfig{height=11cm,file=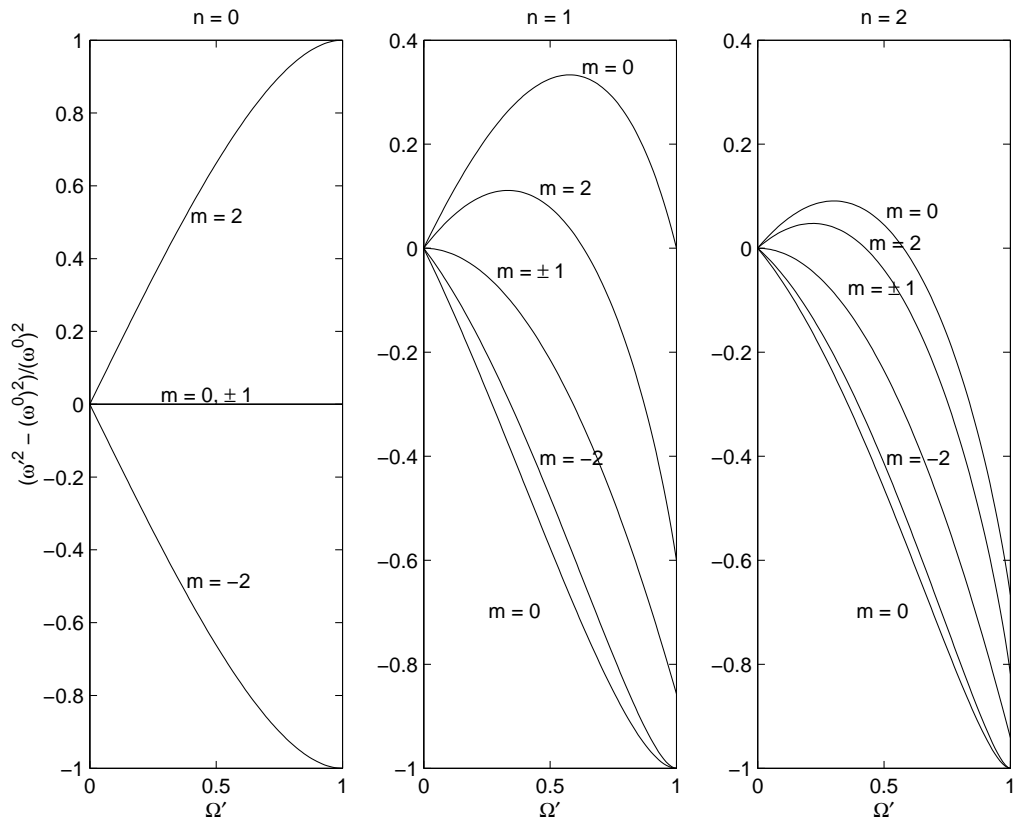}}
\caption{Fractional shift in the squared frequency as a function of the rotation frequency, 
$\Omega'$, for $n = 0,1,2$ and $m = 0, \pm 1, \pm 2$. The three panels give the plots for 
different $n$. The corresponding values of $m$ for each curve are indicated on the figure. 
} \label{fracshift}
\end{figure}

The dispersion curve, $\omega$ vs. $m$, is presented in Fig. \ref{spectrumfig} for 
$\Omega' = 0, 0.25$, $n = 0, 1, 2$, and $m$: $m = 0, \pm 1, \pm 2,  \ldots \pm 10$ in 
the lab frame. 
In addition, it is convenient to define the fractional shift of the squared
frequency: 
\begin{eqnarray}
\frac{\omega^{^{\prime}2}- (\omega^0)^2}{(\omega^0)^2} & =  &  \frac{ \pm 2 (m-1) 
\Omega' 
\sqrt{ \left [ |m|+ 2n(n+|m| + 1) \right ] (1 - \Omega^{^{\prime}2}) + \Omega'^2}}{|m|+ 
2n(n+|m| + 1) } - \Omega^{^{\prime}2}     \nonumber   
\\ 
& + & \frac{  [(m-1)^2 + 1] \Omega'^2  }{|m|+ 2n(n+|m| + 1) } .
\end{eqnarray}
The fractional shift as a function of the rotation frequency in the lab frame, $\Omega'$, for 
$n = 0,1,2$ and $m = 0, \pm 1, \pm 2$ is presented in Fig. \ref{fracshift}.

In summary, we have presented the excitation frequencies of the hydrodynamic shape 
oscillations in a BEC supporting a vortex lattice. 
The methodology we used is similar to that originally applied by Svidzinsky and 
Fetter\cite{FetterNormalMode} for the single vortex excitations; our result may therefore be 
viewed as an
application of their result to the case of vortex lattices. The most pronounced effect of the 
presence of a vortex lattice is clearly the $\Omega$-dependence of the excitation spectrum.

We note that the hydrodynamic amplitudes of the fluctuation in the particle
density and the velocity potential may be written
in terms of the amplitudes $u$ and $v$ of the Bogoliubov
equations, as follows: 
\begin{equation}
n^{\prime}= \Psi^{*} u - \Psi v,
\end{equation}
\begin{equation}
\Phi^{\prime}= \frac{\hbar}{2Mi|\Psi|^2} (\Psi^{*}u + \Psi v),
\end{equation}
and the linearized Euler and the continuity equations may be derived from the equivalent 
GPE\cite{FetterNormalMode}.

The equations derived in this paper are general; the approximations that we have made in
order to model a vortex lattice, and also to solve the equations are in the fluid
velocity field and the TF density profiles of the ground and the excited
states. By refining each of these three approximations one may expect to
improve on the result obtained. For instance, it should be noted that, although 
the TF density is rotationally symmetric, a real vortex lattice state is not 
strictly an eigenstate of $L_z$. In fact, the only vortex 
state which is an angular momentum eigenstate is the case of a single vortex at 
the symmetry axis, and if the single vortex is moved off-axis it gives a value of $L_z$ 
which is not an 
integer\cite{Offcenter}. It is noted that our analytical result Eq. (\ref{spectrum}) still holds 
for non-integer values of $m$ since 
the Jacobi 
Polynomials $P^{(a,b)}_n (x)$ are well defined for all values of $a,b > -1$, 
and  $a \equiv |m|$ and $b \equiv 0$ in our ansatz for $n'$. The excitation spectra 
presented in Fig. \ref{spectrumfig} would therefore be modified accordingly when all such
refinements are taken into account.  As is, our solutions are best applicable to low energy 
excitations of a large condensate with small $\Omega$. 

This work is supported by NSF,ONR, ARO, and University of Rochester. L. O. Baks. is a Horton 
Fellow at L.L.E.

\end{document}